\documentclass[preprint,aps,,showpacs,superscriptaddress]{revtex4}

\usepackage{dcolumn}
\usepackage{graphicx}
\usepackage{amsfonts}
\usepackage{amsmath,bm,amssymb}

\begin{document}

\title{Collinear and non-collinear spin ground state of wurtzite CoO}

\author{Myung Joon Han} 
\email{mj.han@kaist.ac.kr}
\affiliation{Department of Physics, Korea Advanced Institute of
  Science and Technology, Daejeon 305-701, Korea }
\affiliation{KAIST Institute for the NanoCentury, KAIST, Daejeon 305-701, Korea}

\author{Heung-Sik Kim} 
\affiliation{Department of Physics, Korea Advanced Institute of
  Science and Technology, Daejeon 305-701, Korea }

\author{Dong Geun Kim} \affiliation{Department of Physics and
  Astronomy, CSCMR, Seoul National University, Seoul 151-747, Korea}

\author{Jaejun Yu}
\email{jyu@snu.ac.kr}
\affiliation{Department of Physics and
  Astronomy, CSCMR, Seoul National University, Seoul 151-747, Korea}

\date{\today }


\begin{abstract}
  Collinear and non-collinear spin structures of wurtzite phase CoO
  often appearing in nano-sized samples are investigated using
  first-principles density functional theory calculations.  We
  examined the total energy of several different spin configurations,
  electronic structure and the effective magnetic coupling
  strengths. It is shown that the AF3-type antiferromagnetic ordering
  is energetically most stable among possible collinear
  configurations.  Further, we found that a novel spiral spin order
  can be stabilized by including the relativistic spin-orbit coupling
  and the non-collinearity of spin direction. Our result suggests that
  a non-collinear spin ground state can be observed in the
  transition-metal-oxide nanostructures which adds an interesting new
  aspect to the nano-magnetism study.
\end{abstract}

%\begin{keywords}
%CoO nanocrystal, wurtzite structure, non-collinear spin,
%  density functional theory
%\end{keywords}

\maketitle

\section{Introduction}
Recently transition-metal-oxide (TMO) nanocrystals have generated
considerable research interest due to their intriguing material
properties and their potential applications such as storage device and
catalysis
\cite{storage-1,storage-2,storage-3,storage-4,catalysis-1,catalysis-2,
  MnO-APL,MnO-JACS,MnO-Angew,Park-MnO,Han-MnO,e-Fe2O3,WZ-CoO-a,WZ-CoO-b,
  WZ-CoO-An}.  Nano-sized TMO crystals often exhibit different
material characteristics from its bulk counterpart. For example,
unusual magnetic signals have been reported in MnO nanoparticles
\cite{MnO-APL,MnO-JACS,MnO-Angew,Park-MnO}, while bulk MnO is known to
be antiferromagnetic (AFM). The reduced coordination number for the
metal ion at its surface probably affects the electronic structure of
nano-sized TMO \cite{Han-MnO,MnO-Angew}. Furthermore, TMO nanocrystals
often exhibit unconventional structural properties which are not found
in bulk \cite{e-Fe2O3,WZ-CoO-a,WZ-CoO-b,WZ-CoO-An}. Wurtzite (WZ) CoO
is one of the examples. Several experiments independently reported
that WZ phase of CoO is stabilized at the nano-meter scale although
the bulk phase CoO has the rock-salt (RS) structure
\cite{WZ-CoO-a,WZ-CoO-b,WZ-CoO-An}.

The magnetic property of WZ CoO is particularly interesting especially
after room temperature ferromagnetism was found in the Co-doped ZnO
\cite{ZnCoO-roomT-FM-1,ZnCoO-roomT-FM-2,ZnCoO-roomT-FM-3}. The origin
of ferromagnetic signals in the Co-doped ZnO is still under debate,
whereas some experimental efforts have failed to observe the room
temperature ferromagnetism \cite{ZnCoO-fail-1,ZnCoO-fail-2}. Since ZnO
has a WZ structure and doped-Co ions most likely substitue for Zn, the
spin structure of WZ CoO can have important implications for the
magnetic property of Co-doped ZnO.

In spite of previous efforts, the magnetic properties of WZ CoO are
still not clearly understood. An early {\it ab-initio} calculation
based on the local (spin) density approximation (L(S)DA)
\cite{WZ-CoO-b} was followed by a LDA+U (LDA plus Hubbard U) study
which predicts an AFM spin ground state with a finite gap
\cite{Han-JKPS}. By using Monte Carlo simulations, Archer {\it et al.}
also reported AFM spin ordering and estimated the spin exchange
constants as well as the magnetic transition temperatures
\cite{Archer}. In a more recent study, Hanafin {\it et al.}  studied
the shape dependence of the magnetism of WZ CoO nanoparticles by
simulating the Heisenberg spin hamiltonian \cite{Hanafin}. They
performed an extensive Monte Carlo simulation based on exchange
coupling constants obtained from Ref.~\onlinecite{Archer}.
Importantly, however, the ground state spin configuration of the WZ
CoO has not yet been fully explored, and it still remains unclear. For
example, within the collinear spin picture, there may exist several
AFM spin arrangements other than the $c$-type AFM order. Moreover, it
is important in the surface-rich nanocrystals to include the
relativistic spin-orbit couplings.  The inclusion of these effects
combined with AFM coupling can lead to a novel non-collinear spin
structure stabilized in TMO nanocrystals.

\begin{figure}
  \centering \includegraphics[width=17cm]{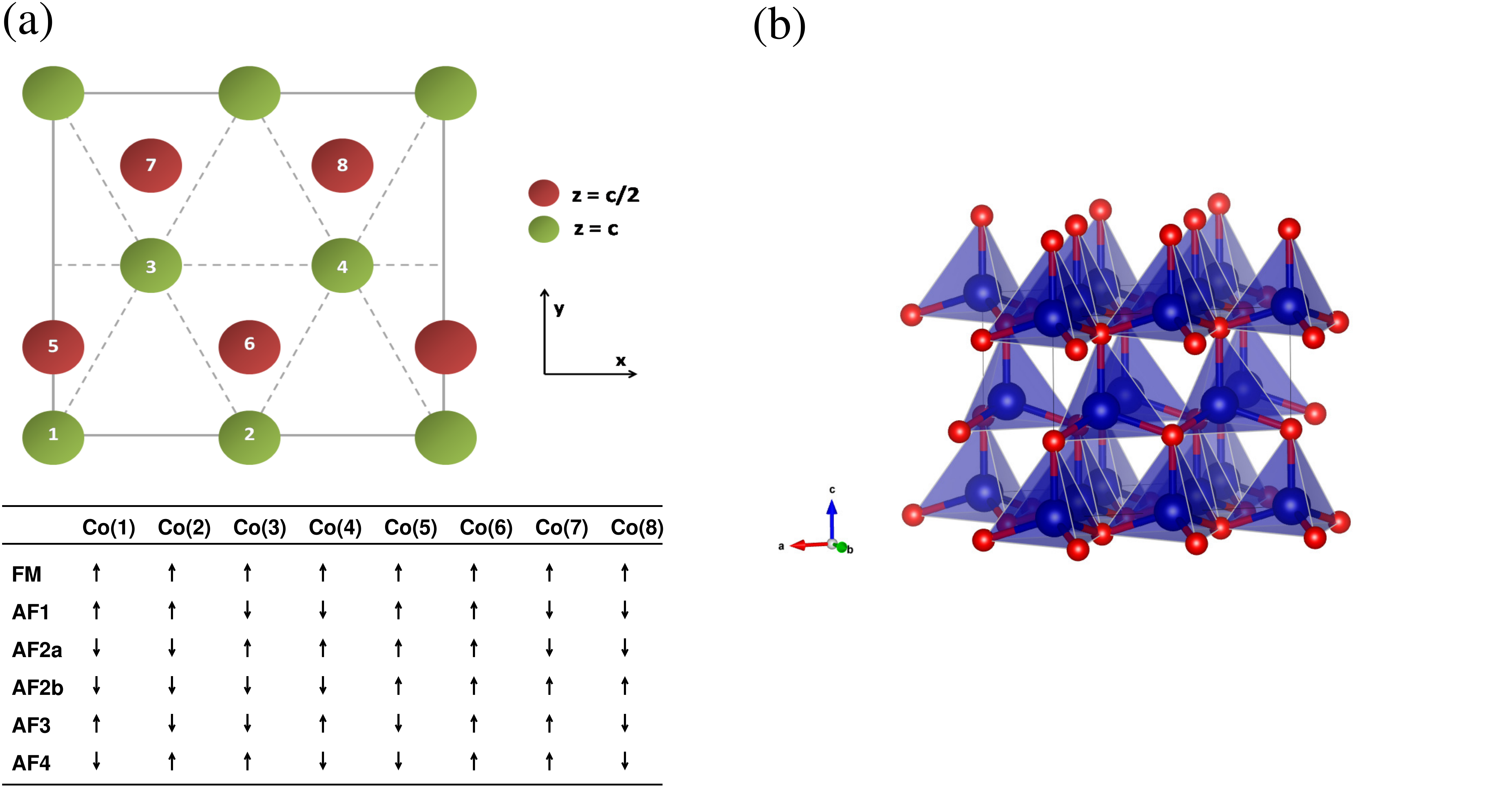}
   \caption{(Color online) (a) Possible spin structures within the
     sixteen-atom unit cell. The upper part: the bright (green) and
     dark (red) circles represent Co atoms at $z=0,c$ and
     $z=\frac{1}{2}c$ plane, respectively. The lower part: up/down
     arrows depict up/down spins, respectively. (b) The ball-and-stick
     figure for the unit cell. The larger and smaller balls represent
     Co and O atoms, respectively.
   \label{fig:unitcell}}
\end{figure}

In this paper, we investigated the magnetic properties of WZ CoO in
detail for the case of non-collinear as well as collinear spin. The
electronic structure and exchange interaction are also investigated.
Within the collinear spin case, one of the AFM spin orderings,
so-called AF3 order, is shown to be most stable energetically, and is
different from $c$-type AFM order. Fully relativistic calculations
show that non-collinear spin configurations may be more stable than
collinear ones. All possible spin configurations were investigated
within sixteen-atom unitcell for the collinear case and
twenty-four-atom unitcell for the non-collinear case. Our results
suggest that a novel spiral spin structure can be realized in CoO
nanocrystals, demonstrating that CoO is a very unique example in the
study of nano-magnetism. That is, the nano-size effect stabilizes a
different structural phase and leads to a novel magnetic ground state
in the nano-meter scale.

\section{Computational Details}

We carried out density functional theory (DFT) calculations for the
periodic unit cell within LDA$+U$ \cite{LDA+U} method by employing a
linear combination of localized pseudo-atomic orbitals method
\cite{Ozaki}.  The effective Coulomb interaction parameter
$U_{eff}=U-J$ is taken as 6 eV, which is proven to be reasonable for
the calculation of RS CoO \cite{Han-LDA+U} while the results of other
$U$ values will also be discussed in the following section.
Ceperley-Alder exchange-correlation energy functional as parameterized
by Perdew and Zunger has been adopted \cite{CA}. Our basis orbitals
were generated by a confinement potential scheme \cite{Ozaki} with
cutoff radii of 4.5 {a.u.} and $5.5$ {a.u.}  for O and Co,
respectively. Troullier-Martins type pseudo-potentials
\cite{troullier} with a partial core correction \cite{Louie} were used
to replace the deep core potentials by norm-conserving soft potentials
in a factorized separable form with multiple projectors proposed by
Bl\"{o}chl \cite{Blochl}. In this pseudo-potential generation, the
semi-core 3$p$ electrons for Co atoms were included as valence
electrons in order to take into account the contribution of the
semi-core states to the electronic structures. Real-space grid
techniques \cite{SIESTA} were used with an energy cutoff of 220 Ry in
numerical integrations. In addition, the projector expansion method
was employed to accurately calculate three-center integrals associated
with a deep neutral atom potential \cite{Ozaki2}.  For lattice
parameters, we consider the experimental values of a WZ CoO nanorod,
prepared by the thermal decomposition of a cobalt-oleate complex:
$a=3.249$ \AA~ and $c=5.206$ \AA~\cite{WZ-CoO-An}, which are
consistent with the values reported by Seo {\it et
  al.}~\cite{WZ-CoO-a}.  For non-collinear calculations, we generated
$j$-dependent pseudopotential, by solving the Dirac equation instead
of the conventional Schr{\"o}dinger equation, in which the fully
relativistic effect as well as the spin-orbit coupling terms were
included ~\cite{NCDFT}.  In this computation scheme, the spins are
represented by a spinor matrix, and therefore, the angles between Co
spins can have arbitrary values \cite{OpenMX}.

\section{Result and Discussion}

\begin{table}[t]
\begin{ruledtabular}                                 
\begin{tabular}{cccccc}
      & AFM (in) & FM (in) & AFM (out) & FM (out) & E$_{\rm tot}$\\
\hline
 FM   &  0  &  6  &  0  &  6  & 0.072 \\
 AF1  &  4  &  2  &  4  &  2  & 0.002 \\
 AF2a &  4  &  2  &  2  &  4  & 0.010 \\
 AF2b &  0  &  6  &  6  &  0  & 0.041 \\
 AF3  &  4  &  2  &  4  &  2  & 0.000 \\
 AF4  &  4  &  2  &  2  &  4  & 0.009 \\
\end{tabular}   
\end{ruledtabular}
\caption{\label{tab:CL-E} Total number of AFM and FM couplings and the
  calculated total energies for six different collinear spin
  configurations (in the unit of eV/CoO). The terms `in' and `out'
  refer to the number of nearest couplings along the `in-plane' and
  `out-of-plane' directions, respectively. The energy of the most
  stable AF3 is set to be zero.  }
\end{table}

\begin{figure}
  \centering \includegraphics[width=8cm]{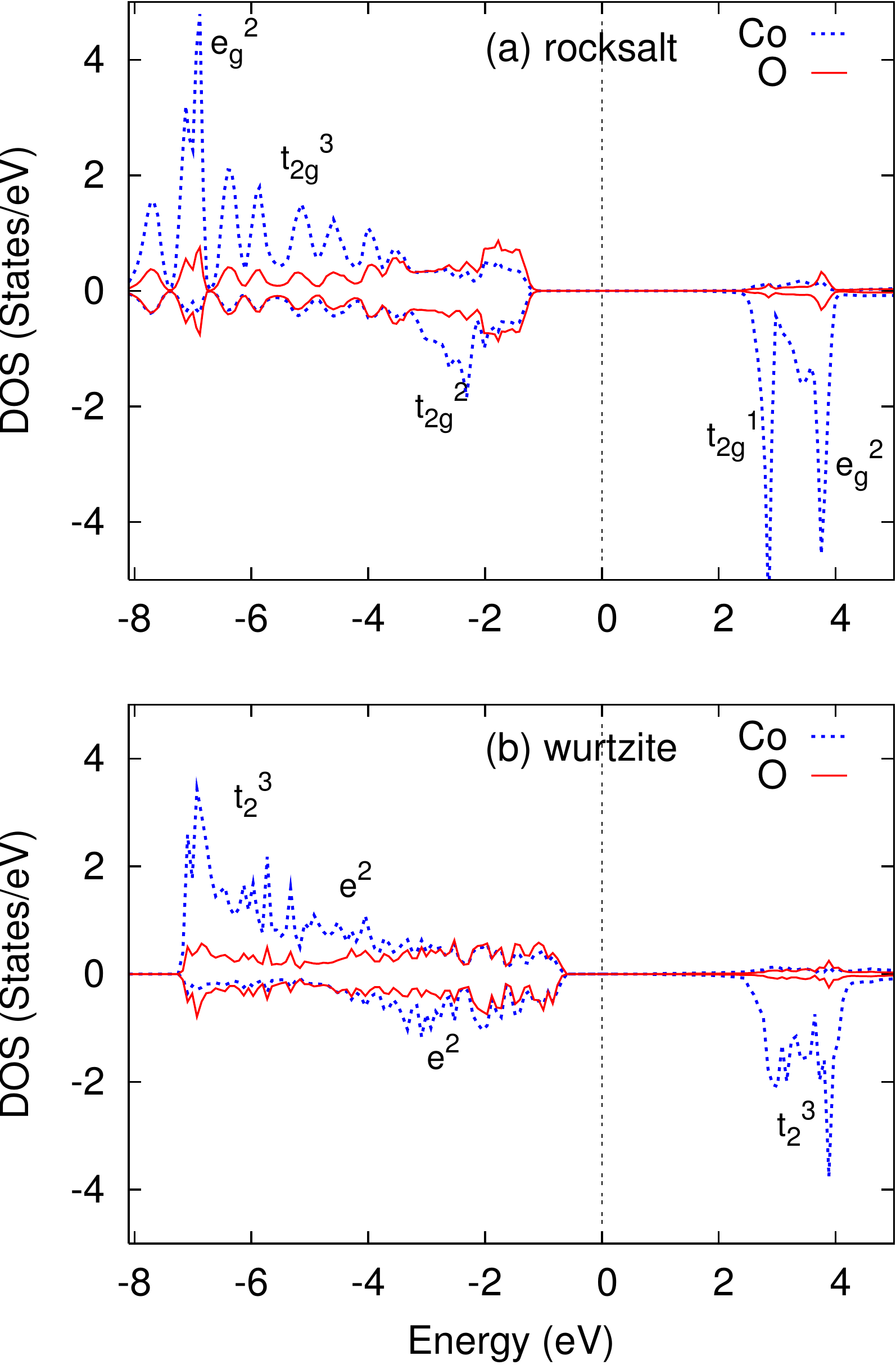}
   \caption{(Color online) The projected-DOS of (a) RS CoO and (b) AF3
     phase of WZ CoO. Dotted (blue) and solid (red) lines represent
     the states projected onto the Co and O sites, respectively. The
     Fermi level is set to be zero (vertical dotted
     line). \label{fig:PDOS-WZ-AF3}}
\end{figure}

\begin{figure*}[t]
\begin{center}
\includegraphics[width=0.6\columnwidth, angle=0]{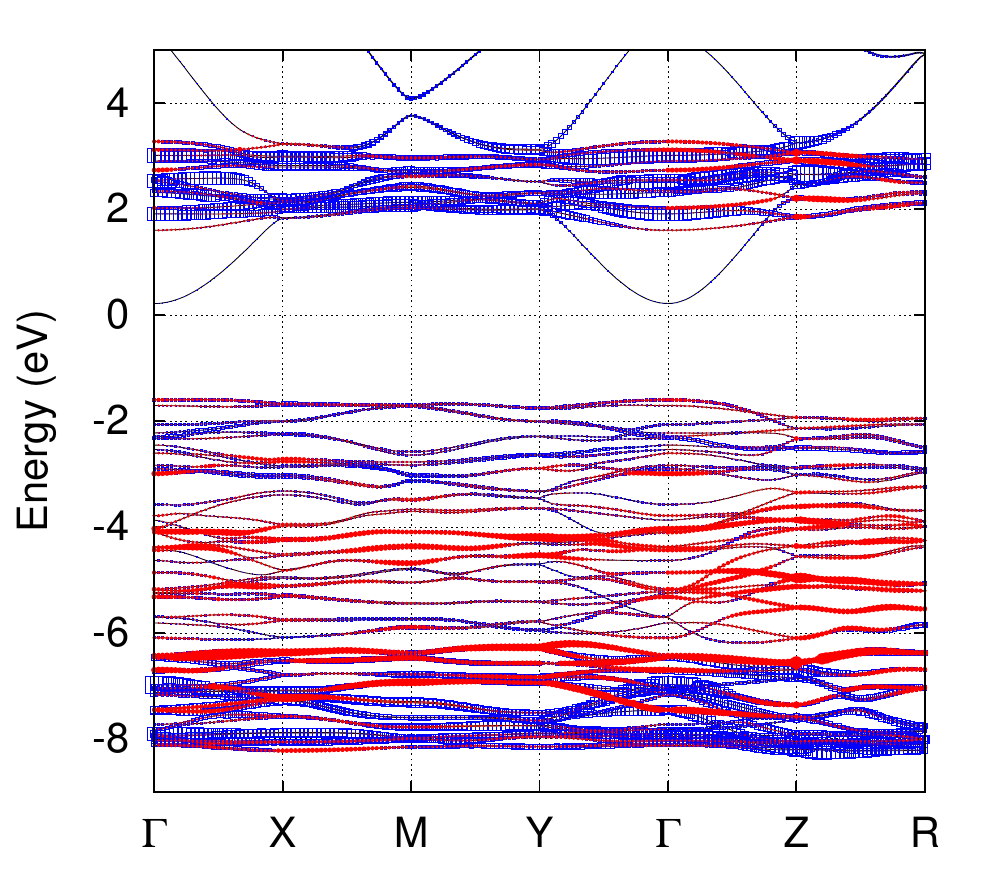}
\includegraphics[width=0.6\columnwidth, angle=0]{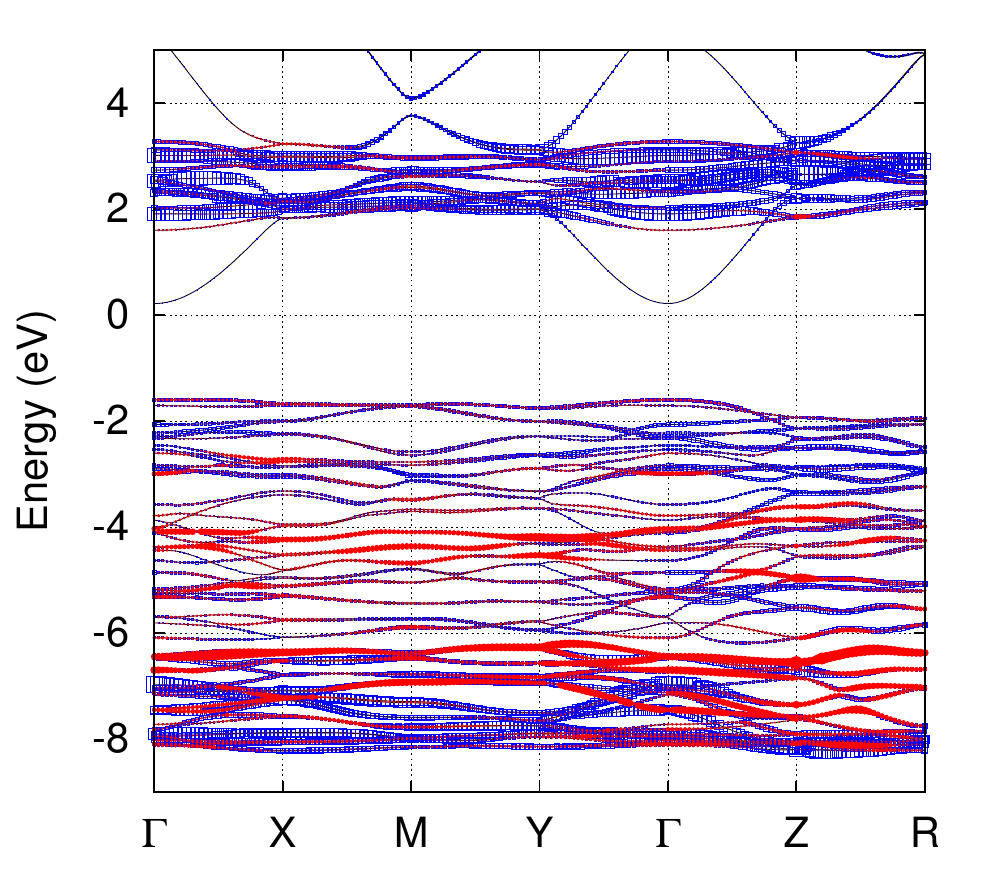}
\caption{(Color online) The calculated fat-band dispersion. The upper
  and lower panel represent the up and down spin bands,
  respectively. The blue-boxes and red-circles represent the $t_{2g}$
  ($t_2$) and $e_{g}$ ($e$) components with the symbol size
  proportional to the weight of each component.
\label{fig:band}}
\end{center}
\end{figure*}

%\subsection{Collinear spin ground state}

Figure~\ref{fig:unitcell}(a) shows the six possible collinear spin
structures within the sixteen-atom unitcell adapted in this study
where the dark (red) and bright (green) spheres represent Co atoms at
$z=c/2$ and $z=c$, respectively. The up/down spins are depicted by
up/down arrows in the lower part of Fig.~\ref{fig:unitcell}(a). The
ball-and-stick figure for the unitcell is presented in
Fig.~\ref{fig:unitcell}(b).  In addition to the ferromagnetic (FM)
order, five different AFM configurations can be considered, namely
AF1, AF2a, AF2b, AF3, and AF4 \cite{MnS}. The calculated total
energies (E$_{\rm tot}$) are presented in Table~\ref{tab:CL-E} where
E$_{\rm tot}$ of AF3 is set to be zero. The results show that AF3 is
stabler than any other AFM phases by 0.002--0.041 eV/CoO and than FM
by $0.072$ eV/CoO. This order of stability among the spin
configurations is found to be robust against $U$-value. The
calculations with $U$=4 and 8 eV show that the deviations in the
calculated E$_{\rm tot}$ relative to AF3 can be different by
$\sim$15--25\% and the order of E$_{\rm tot}$ remains same with $U$=6
eV result.  Since the ground state spin configuration is AFM, our
result implies that the magnetic signal previously detected by X-ray
magnetic circular dichroism (XMCD) measurement \cite{WZ-CoO-An} is not
attributed to WZ phase of CoO although the contribution from the
uncompensated surface moments cannot be ruled out \cite{Archer}. Also,
the room temperature ferromagnetism observed in Co-doped ZnO is not
likely an intrinsic effect from the ferromagnetic ordering of Co
spins, as concluded by recent Monte Carlo simulations \cite{Hanafin}.

Note that AF2b corresponds to the $c$-type AFM order that was studied
by Risbud {\it et al.} within LSDA ($U=0$) \cite{WZ-CoO-b} and by two
of the authors of this paper within LDA$+U$ \cite{Han-JKPS}. While
AF2b is more stable than FM \cite{Han-JKPS}, its total energy is
notably higher than the other AFM configurations. The $c$-type AFM
order is therefore not the ground state configuration (even among the
collinear spin structures) and cannot be realized in the experimental
situation.  To verify the other possibility, we also performed total
energy calculations for the several uncompensated AFM configurations
that carry the net moments. Note that these are rather artificial spin
orders because they are not commensurate with the unitcell. They are
found to have significantly higher energies compared to the most
stable AF3 by 0.018--0.022 eV/CoO.  Thus, such unnatural spin orders
carrying net moments can also be ruled out as the source of the
observed moment Ref.~\onlinecite{WZ-CoO-An}.

The relative stability of these collinear spin configurations can be
understood by counting the number of AFM and FM couplings, as
summarized in Table~\ref{tab:CL-E}. Since the bond angles between Co
ions are approximately 110$^{\circ}$, AFM couplings are probably
favored. Thus, the configuration that maximizes the number of AFM
pairs (and minimizes FM pairs), would become most stable.  As shown in
Table~\ref{tab:CL-E}, this is the case for AF1 and AF3. AF2a and AF4
have higher energies than AF1 and AF3 by 7--10 meV/CoO due to their
fewer AFM (and more FM) couplings. Although AF2b has the same number
of AFM pairs as AF2a and AF4, all of its AFM pairs are along the
out-of-plane direction. Thus, its higher total energy reflects the
different in-plane magnetic interactions from the out-of-plane
ones. Our simple rule of nearest-neighbor number counting is further
confirmed by the fact that AF1 and AF2a have almost same energy as AF3
and AF4, respectively; they actually have the same number of
nearest-neighbor FM/AFM couplings. The small energy differences,
$\sim$0.001 eV/CoO, can reflect the effect from the longer range
interactions.

From the calculated total energies, we estimated the exchange coupling
parameter in the Heisenberg spin Hamiltonian, $H_{ij}=-J\sum S_i
S_j$. Considering the number of AFM and FM couplings shown in Table
\ref{tab:CL-E}, the in-plane and out-of-plane interactions as can be
calculated follows:
\begin{equation}
J_{\rm in}^{\rm WZ} = \frac{1}{4} (E^{1,3}_{\rm tot} 
                     - \frac{1}{2}E^{2a,4}_{\rm tot}) 
                     = -3 ~{\rm meV},
\end{equation}
\begin{equation}
J_{\rm out}^{\rm WZ} = \frac{1}{4} (E^{1,3}_{\rm tot} 
                      -\frac{1}{12}E^{2b}_{\rm tot}) 
                      = -1 ~{\rm meV}.
\end{equation}
In these equations, $E^{m}_{\rm tot}$ represents the calculated total
energy (per CoO) for the spin configuration (denoted by $m$), and we
used $S=\pm 1.5$ for Co spin which is in good agreement with the
calculated moment. The estimated AFM interactions for both in-plane
and out-of-plane spin directions are consistent with the total energy
results shown in Table \ref{tab:CL-E}: {\it e.g.,} AF2b phase, is less
stable than AF2a and AF4 because $|J_{\rm in}| < |J_{\rm out}|$ (see,
Fig.~\ref{fig:unitcell} and Table~\ref{tab:CL-E}).  Note that the
nearest-neighbor couplings are AFM, which can be supported also by the
Monte Carlo calculations \cite{Archer,Hanafin}.  The difference
between the coupling strengths in Ref.~\onlinecite{Archer} and our
results reflects the different spin structures assumed in the two
studies.  It is useful to compare our result with the hypothetical WZ
MnO case reported by Gopal {\it et al.}~\cite{WZ-MnO}: The in-plane
and out-of-plane couplings for WZ MnO are $J_{\rm in}^{\textrm{MnO}} =
-3.7$ meV and $J_{\rm out}^{\textrm{MnO}} = -4.5$ meV, respectively
\cite{comment2}. It is noted that they are also AFM along both
directions and with the same order of magnitude, whereas in MnO the
out-of-plane coupling is stronger than the in-plane one. It should be
noted that the $J$ values in WZ-MnO were estimated based on the LSDA
($U=0$) total energies, which can produce a substantial difference
from LDA$+U$ results.

The electronic structure of WZ CoO is different from conventional RS
CoO owing to its tetrahedral crystal field. Since Co$^{2+}$ ions are
located in the oxygen tetrahedra instead of the octahedra, their $3d$
levels are split into low-lying $e$ and higher $t_2$ bands
\cite{WZ-CoO-b,Han-JKPS}, with seven electrons occupying the up-spin
$e^{\uparrow}$ and $t_2^{\uparrow}$ bands and the down-spin
$e^{\downarrow}$ bands.  It is consistent with the calculated Co
magnetic moment $\sim$3$\mu_B$.  The projected densities-of-states
(PDOS) for the most stable AF3 are shown and compared to those in RS
phase in Fig.~\ref{fig:PDOS-WZ-AF3}. The band dispersion is presented
in Fig.~\ref{fig:band} where the orbital characters are represented
by shape and size of the symbols.  While the band structure is quite
similar to that of the $c$-type AFM phase obtained by LDA$+U$ in
Ref.~\onlinecite{Han-JKPS}, it is clearly different from the LSDA
($U$=0) results in Ref.~\onlinecite{WZ-CoO-b}.  The finite gap and
large exchange splitting are attributed to the on-site correlation,
$U$, which cannot be well captured by the LSDA ($U=0$)
\cite{WZ-CoO-b}.  Also, the LDA$+U$ valence bands are dominated by the
O-$2p$ and Co-$3d$ mixture whereas the LSDA bands are of mainly Co
character \cite{WZ-CoO-b}.

%\subsection{Non-collinear spin ground state}

\begin{figure}
  \centering \includegraphics[width=9cm]{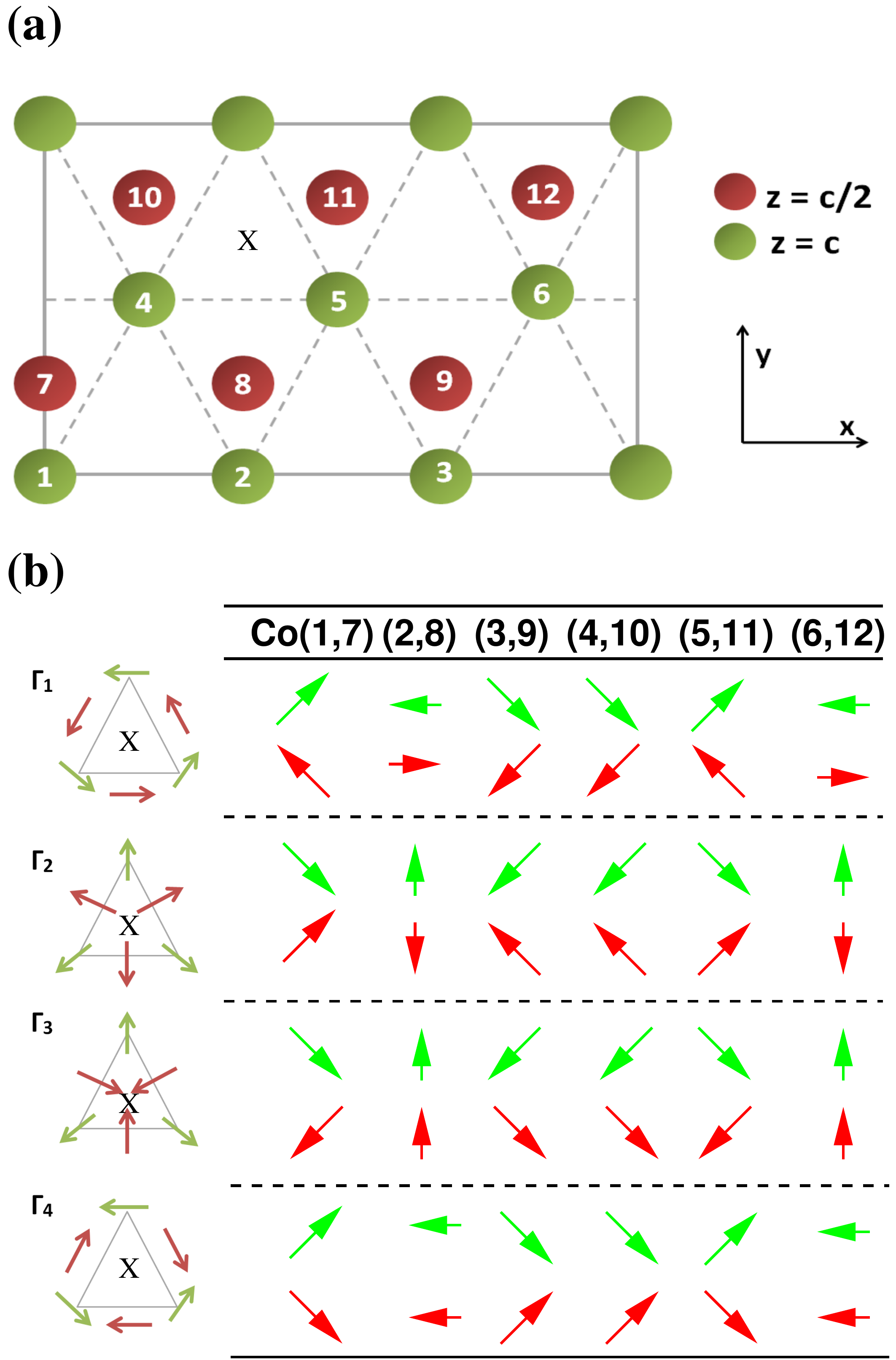}
  \caption{(Color online) Unicell spin structures for non-collinear
    configurations. Bright (green) and dark (red) circles in (a) and
    arrows in (b) represent the Co atoms at $z=0,c$ and
    $z=\frac{1}{2}c$ plane, respectively. The arrows in (b) indicate
    the spin direction within the $ab$-plane.
    \label{fig:NC-uc}}
\end{figure}

Finally, we note that in WZ structure, Co spins form hexagonal
networks and their exchange couplings are all AFM as discussed above.
It suggests that collinear spin arrangements may not be well
stabilized, but a non-collinear order is realized. For the further
examination of magnetic ground states, the relativistic calculations
including spin-orbit couplings were performed. We constructed
commensurate non-collinear spin structures by enlarging our unitcell
to contain 24 atoms (see Fig.~\ref{fig:NC-uc}).  All spins are assumed
to align in the $ab$-plane, which is supported by a recent Monte Carlo
calculation \cite{Hanafin}. Within this unitcell, four different
non-collinear configurations can be considered, namely
$\Gamma_1$--$\Gamma_4$ (see Fig.~\ref{fig:NC-uc}).  The configurations
are distinguishable by their topologies, as shown in
Fig.~\ref{fig:NC-uc}(b), while all the angles between in-plane spins
are $120^{\circ}$. Note that all the configurations,
$\Gamma_1$--$\Gamma_4$, carry zero total moments.

Total energy calculations show that the non-collinear spin structures
are more stable than the collinear configurations.  The collinear spin
orders have higher energies than the non-collinear configurations by
more than 21 meV although this energy difference can be as small as
$\sim\pm$0.1 meV/CoO depending on the details of computation such as
the anisotropy of spin direction, spin-orbit coupling, and the
constraints in the non-collinear calculations. Therefore, our results
demonstrate that a novel non-collinear spin ordering can be realized
in CoO nanostructures.  $\Gamma_3$ is found to be most stable among
the non-collinear configurations, whereas $\Gamma_1$ has almost same
energy as $\Gamma_3$; the difference between the two is
E$(\Gamma_1-\Gamma_3)=0.02 ~{\rm meV/CoO}$.  $\Gamma_2$ and $\Gamma_4$
have higher energies than $\Gamma_1$ and $\Gamma_3$ by 7.28 and 7.34
meV/CoO, respectively.

This type of non-collinear spin order has not been found in TMO
nanostructures. We also emphasize that this novel spin ground state
becomes stabilized through the structural transformation caused by a
material-size reduction into the nano-meter scale, which can add a new
interesting aspect to nano-magnetism studies. We hope that this work
stimulates further research efforts from both experimental and
theoretical perspectives.

\section{Conclusion} Magnetic properties of WZ structure CoO often
appearing in CoO nanocrystals have been studied by using LDA$+U$
density functional method.  Total energy calculations show that the
novel non-collinear spin order can be stabilized, whereas so-called
AF3 type is the most stable among the collinear spin
configurations. Non-collinear spin structures stabilized in the
nano-meter scale are expected to provide a new aspect to
nano-magnetism studies.

\section{acknowledgments}
We are grateful to Prof.~T.~Ozaki and Dr.~B.-J.Yang for
discussions. This work was supported by the NRF through ARP (Grant
No. R17-2008-033-01000-0). This work
was supported by the National Institute of Supercomputing and
Networking / Korea Institue of Science and Technology Information with
supercomputing resources including technical support
(KSC-2013-C2-005).

\end{document}